# Comparison of Neural Models for X-ray Image Classification in COVID-19 Detection

Jimi Togni, Romis Attux
DSPCOM / FEEC / UNICAMP - Brazil
jimitogni@gmail.com, attux@dca.fee.unicamp.br

**Abstract**

This study presents a comparative analysis of methods for detecting COVID-19 infection in radiographic images. The images, sourced from publicly available datasets, were categorized into three classes: 'normal,' 'pneumonia,' and 'COVID.' For the experiments, transfer learning was employed using eight pre-trained networks: SqueezeNet, DenseNet, ResNet, AlexNet, VGG, GoogleNet, ShuffleNet, and MobileNet. DenseNet achieved the highest accuracy of 97.64% using the ADAM optimization function in the multiclass approach. In the binary classification approach, the highest precision was 99.98%, obtained by the VGG, ResNet, and MobileNet networks. A comparative evaluation was also conducted using heat maps.

## I. Introduction

In 2020, the world was profoundly affected by a pandemic caused by a virus known as SARS-CoV-2 (commonly referred to as 'coronavirus' [1]). After a rapid surge in cases in China, the disease spread globally, with over 150 million confirmed cases and 3.2 million deaths reported by early May 2021.

According to Brazil's Ministry of Health [2], the two most widely used detection methods are real-time RT-PCR and partial or complete sequencing of the viral genome. Another technique involves Rapid Testing, which detects antibodies (IgM and IgG) against the coronavirus in infected individuals, based on lateral flow chromatographic methodology. While these tests are essential, limitations such as processing time and cost create opportunities for automated detection tools.

Numerous studies address COVID-19 detection using X-ray imaging, as highlighted in recent surveys by S. Bhattacharya et al. (2021) [3]. These works significantly informed our research. For example, S. Basu (2020) [4] employed transfer learning in a context involving four classes: 'normal,' 'pneumonia,' 'COVID-19,' and 'other diseases,' achieving an overall accuracy of 90.13%. Similarly, R. Jain et al. (2021) [5] applied pre-trained models such as Inception V3, Xception, and ResNeXt to a dataset of 6,432 chest X-ray images, with 5,467 for training and 965 for validation. The Xception network achieved the highest accuracy of 97.97%.

This study aims to conduct a comparative analysis of methods for detecting conditions caused by SARS-CoV-2 through chest X-ray images. We propose using Convolutional Neural Networks (CNNs) and, given the relatively limited data available, adopting transfer learning (TL). The chest X-ray images, obtained from public datasets [6], are classified into three categories: 'normal' (no COVID-19 characteristics), 'pneumonia' (features indicative of pneumonia), and 'COVID' (patterns of SARS-CoV-2 infection). The CNNs were trained on images pre-labeled by specialists in a supervised approach. Infected patients often experience severe pneumonia, pulmonary edema, acute respiratory distress syndrome (ARDS), or multiple organ failure, leading to distinct patterns in their X-ray images [1].

The structure of this paper is as follows: Section II outlines the study's objectives and proposal, Section III details the methodologies, Section IV examines the image features that differentiate the classes using heatmaps, Section V discusses the results, and Section VI concludes the findings.

## II. Proposal

This study employs an online approach for dataset generation. In this approach, the physical dataset remains unaltered: transformations are applied incrementally during the training epochs of the model. That is, with the conclusion of each training epoch, the images are transformed according to predefined specifications. This strategy, commonly used in the literature, is particularly important due to the relatively small number of available images, which poses a risk of bias.

For the multiclass classification tests, radiographic images labeled by specialists were used in three classes: 'normal,' 'pneumonia,' and 'COVID.' In the binary tests, only the 'pneumonia' and 'COVID' classes were considered. A total of 3,856 images were obtained, sourced from the Kaggle website [6]. The original images lacked standardization regarding size, resolution, or brightness; this was addressed during preprocessing by resizing all images to a uniform size. During the online data augmentation tests, all images underwent the following transformations: resizing to 224x224 pixels, random rotation between -10 and +10 degrees around a fixed central point, and horizontal flipping with a 5% probability per image per epoch. Figure 1 displays random examples of transformed images from each class. The final number of images in the training, validation, and test datasets is shown in Tables I and II.

## III. Evaluated Methodologies

A key aspect of this study was the application of transfer learning (TL) [7]. Transfer learning enables the use of a set of layers from a pre-trained CNN. In tasks like the one addressed in this research, one possible approach is to utilize networks that have been tested in image recognition competitions, such as the ImageNet Large Scale Visual Recognition Challenge (ILSVRC) or the ImageNet Challenge [8]. These models are trained on extensive datasets comprising various categories, such as animals, vehicles, and everyday objects [9]. Both the datasets and some pre-trained networks are publicly available, facilitating the creation and evaluation of new models.

In TL, it is possible to use pre-trained networks and further train specific parts of their architecture. This means that the initial layers—trained for general image recognition tasks—can remain unchanged, while only the final layers are fine-tuned for the specific problem at hand [10]. This is the approach adopted in this study, motivated by the limited availability of chest X-ray images for training a classification system involving COVID-19.

The dataset was divided into three parts: Training, Validation, and Testing. Initially, the division was 80% for training, 19% for validation, and 1% for testing. This allocation, with only 1% for testing, was chosen because new chest X-ray images classified as 'COVID' were being added daily to the GitHub repository [11] at the time of the study. Including these new images in the training and validation datasets was impractical, as it would require retraining the models. Therefore, all new images received after training were added exclusively to the test dataset, ensuring that the models were evaluated for their ability to generalize to new images. Tables I and II shows the total number of images and their distribution across the datasets.

For all tests, the number of epochs (EP) was set to 100, and the learning rates (LR) were $10^{-3}$, $10^{-4}$, and $10^{-5}$. A diverse and representative set of optimizers (OPT) was chosen: ADAM (Adaptive Moment Estimation), SGD (Stochastic Gradient Descent), and RMSProp [12]. The batch size (BS), representing the number of samples processed per training iteration, varied between 8, 16, and 32. Cross-Entropy (CE) was used as the loss criterion for multiclass classification, while Binary Cross-Entropy (BCE) was used for binary classification. Accuracy (ACC) is indicated in the tables. The PyTorch framework facilitated the implementation of the machine learning techniques tested.

Table I
Total number of images per class.

| Class | Images |
|---|---|
| Normal (Nm) | 1,341 |
| Pneumonia (Pn) | 1,345 |
| COVID (Cv) | 1,170 |
| **Total** | **3,856** |

Table II
Number of images per class divided across training, validation, and test datasets.

| Class | Training | Test | Validation |
|---|---|---|---|
| Normal (Nm) | 1,051 | 14 | 255 |
| Pneumonia (Pn) | 1,097 | 13 | 240 |
| COVID (Cv) | 937 | 11 | 238 |
| **Total** | **3,085** | **38** | **733** |

Based on the literature and preliminary tests, the following networks were selected to form the classifiers through TL: AlexNet [13], the winner of the ImageNet 2012 competition [8]; DenseNet [14], known for its architecture promoting maximum information flow between layers; GoogleNet [15], which used 12 times fewer parameters than the winning network of the ILSVRC 2014 competition; MobileNet [16], designed for mobile and resource-constrained environments, reducing computational and memory requirements while maintaining accuracy; ResNet [17], a 152-layer residual network that won the ILSVRC 2015 classification task; ShuffleNet [18], which achieved superior performance at 40 MFLOPs; SqueezeNet [19], achieving AlexNet-level accuracy on ImageNet with 50 times fewer parameters; and VGG [20], which ranked first and second in the ImageNet Challenge 2014 in object detection and classification, respectively [21].

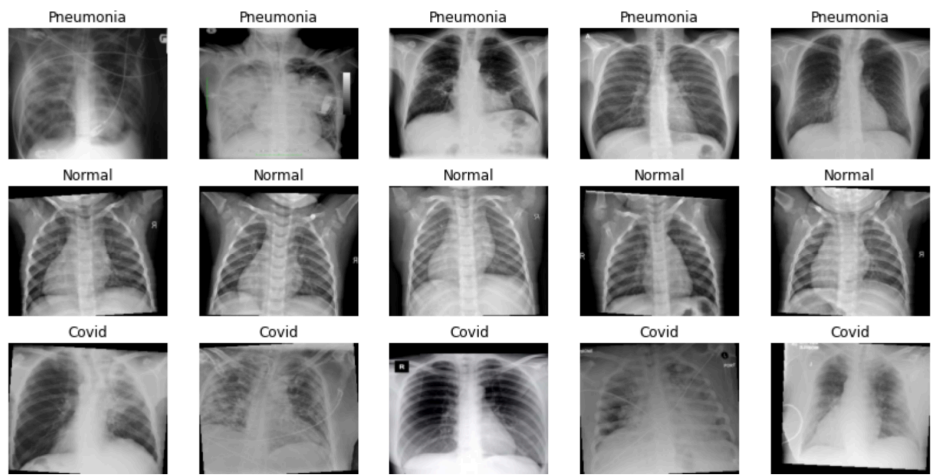

**Figure 1.** Sample images for each class with the transformations applied during data augmentation.

**IV. Heat Maps**

Class Activation Mapping (CAM), also referred to as heat maps, aims to identify the regions within an image that are particularly significant for the network during the classification process. Red regions indicate strong activation, yellow denotes moderate activation, and light blue to dark blue represents low or no activation.

Figures 2 and 3 illustrate that, from the perspective of the convolutional neural networks utilized in this study, the 'COVID' and 'pneumonia' classes exhibit marked differences. For 'pneumonia,' the regions of highest activation (red areas) are concentrated on the left side of the thorax. This pattern appears consistent with explanations provided on the "Radiology Assistant" website—a resource by the Radiological Society of the Netherlands [22]. However, a more rigorous analysis would require the involvement of a specialist in future studies. Conversely, images classified as 'COVID' exhibit strong activation in the upper right and left regions of the thorax, consistent with observations discussed in [23].

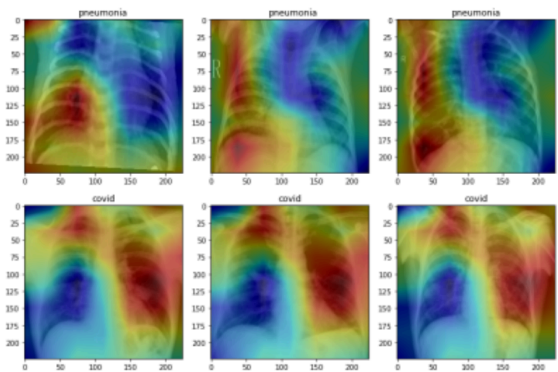

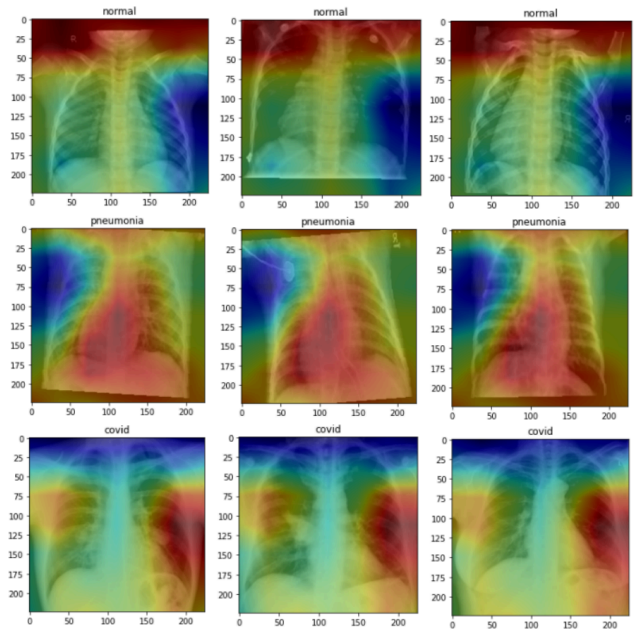

**Figure 2.** Heat map for multiclass classification.

**Figure 3.** Heat map for binary classification.

## V. Results

Tables III and IV present the best results obtained for each network along with the corresponding optimal hyperparameters. Sensitivity, precision, and F1 score values for the multiclass approach are displayed in Table V, while the binary classification approach is summarized in Table IV.

**Table III**
Hyperparameters leading to the best validation accuracy for the **multiclass** approach.

| Model | EP | LR | BS | OPT | CRT | ACC |
|---|---|---|---|---|---|---|
| AlexNet | 100 | $10^{-4}$ | 8 | ADAM | Cross Entropy | 94.50% |
| DenseNet | | | 16 | | | 97.64% |
| GoogleNet | | | 8 | | | 95.81% |
| MobileNet | | | 8 | | | 96.46% |
| ResNet | | | 8 | | | 96.20% |
| ShuffleNet | | | 8 | | | 95.15% |
| SqueezeNet | | | 16 | | | 96.60% |
| VGG | | | 8 | | | 96.20% |

**Table IV**
Hyperparameters leading to the best validation accuracy for the **binary** approach.

| Model | EP | LR | BS | OPT | CRT | ACC |
|---|---|---|---|---|---|---|
| AlexNet | 100 | $10^{-4}$ | 16 | SGD | Binary Cross Entropy | 99.06% |
| DenseNet | | | | ADAM | | 99.94% |
| GoogleNet | | | | ADAM | | 99.63% |
| MobileNet | | | | ADAM | | 99.98% |
| ResNet | | | | ADAM | | 99.98% |
| ShuffleNet | | | | RMSprop | | 99.81% |
| SqueezeNet | | | | SGD | | 99.56% |
| VGG | | | | RMSprop | | 99.98% |

Table VII outlines the performance of each network for all classifiers tested, aiming to identify the approach yielding the highest overall accuracy. Following tests with all hyperparameters described in Section III, the best result was achieved with a learning rate of $10^{-4}$ in combination with the ADAM optimizer, as shown in Table III. Nonetheless, other optimizers such as SGD and RMSprop demonstrated comparable or superior performance in certain cases, for example, with the MobileNet and ResNet models, as indicated in Table VII.

**Table V**
Highest precision achieved per class for each network in the **multiclass** approach.

| Model | Sensitivity (%) | Precision (%) | F1 Score (%) |
|---|---|---|---|
| Classes | Normal | Covid | Pneumonia |
| AlexNet | 96 | 95 | 91 |
| DenseNet | 97 | 97 | 96 |
| GoogleNet | 97 | 97 | 96 |
| MobileNet | 96 | 95 | 95 |
| ResNet | 95 | 94 | 95 |
| ShuffleNet | 96 | 94 | 93 |
| SqueezeNet | 97 | 97 | 95 |
| VGG | 94 | 95 | 95 |

**Table VI**
Highest precision achieved per class for each network in the **binary** approach.

| Model | Sensitivity (%) | Precision (%) | F1 Score (%) |
|---|---|---|---|
| Class | Normal | Covid | Pneumonia |
| AlexNet | 99 | 99 | 99 |
| DenseNet | 99 | 99 | 99 |
| GoogleNet | 100 | 99 | 99 |
| MobileNet | 100 | 100 | 100 |
| ResNet | 100 | 100 | 100 |
| ShuffleNet | 100 | 100 | 100 |
| SqueezeNet | 100 | 100 | 100 |
| VGG | 99 | 99 | 99 |

The highest result across all networks tested for the multiclass approach was achieved by the DenseNet architecture, as demonstrated in Table VII. With the ADAM optimizer, an accuracy of 97.64% was obtained. Sensitivity, precision, and F1 scores for DenseNet can also be observed in Table V. Figure 4 illustrates the confusion matrices for DenseNet under both multiclass (Figure 4, left) and binary (Figure 4, right) configurations.

**Table VII**
Best accuracy achieved for each classifier and tested optimizers in the **multiclass and binary** approach.

| Model | Average Accuracy (%) | |
|---|---|---|
| Approach | Multiclass (n, %) | Binary (n, %) |

| Optimizer | SGD | ADAM | RMSProp | SGD | ADAM | RMSProp |
|---|---|---|---|---|---|---|
| SqueezeNet | 95.94% | 96.59% | 96.33% | 99.56% | 99.08% | 98.08% |
| DenseNet | 97.25% | 97.64% | 97.38% | 99.90% | 99.94% | 99.05% |
| ResNet | 96.22% | 96.01% | 96.18% | 99.10% | 99.63% | 99.98% |
| VGG | 97.38% | 97.59% | 97.55% | 99.52% | 99.77% | 99.98% |
| AlexNet | 95.85% | 95.76% | 95.76% | 99.06% | 98.88% | 98.76% |
| GoogleNet | 96.53% | 96.55% | 96.43% | 99.63% | 99.23% | 98.87% |
| MobileNet | 97.51% | 97.51% | 97.47% | 99.83% | 99.98% | 99.87% |
| ShuffleNet | 92.32% | 97.18% | 96.93% | 99.55% | 99.53% | 99.81% |

Figure 5 shows the evolution of accuracy and error rate for the multiclass problem, while Figure 6 depicts the results for the binary classification approach.

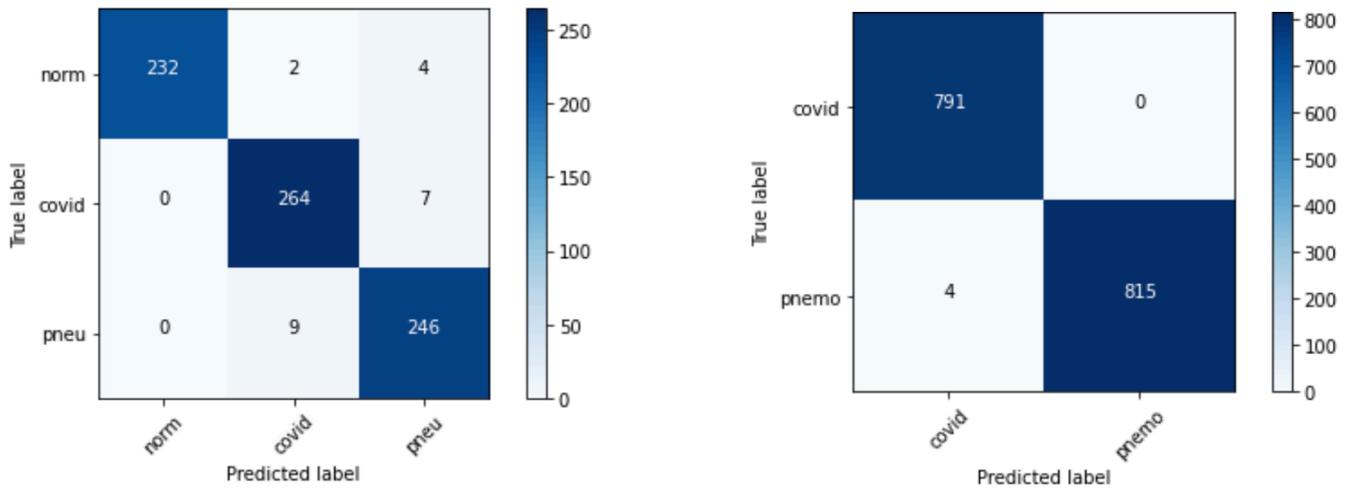

**Figure 4**. Confusion matrix for the multiclass approach on the left and results of the binary approach on the right

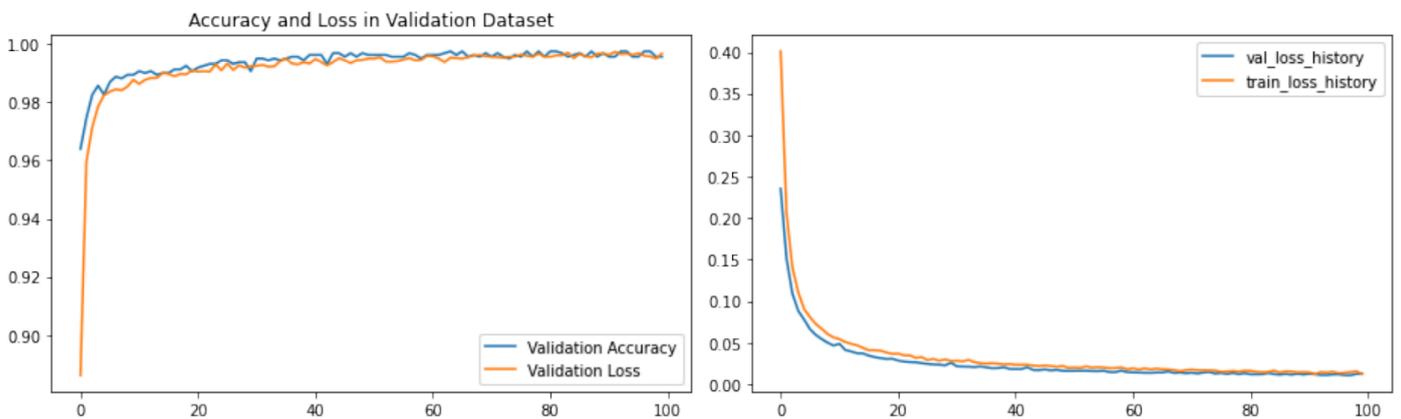

**Figure 5.** Curves for the multiclass approach showing the best result obtained among all tested networks using the DenseNet model.

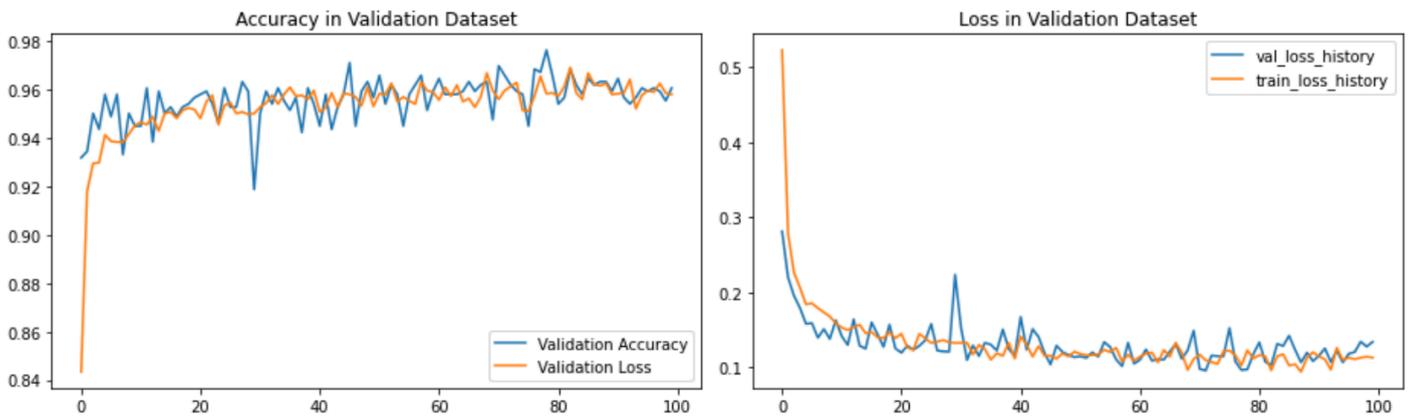

**Figure 6.** Curves for the binary approach showing the best result obtained with the MobileNet model, which achieved the highest accuracy among the three tested optimizers, as shown in Table VIII.

## VI. Conclusions

This study conducted a comparative analysis of different neural network approaches to solve the problem of classifying chest X-ray images. The objective was to facilitate the detection of patterns corresponding to those of COVID-19 patients, contributing to research in this critically relevant field.

The findings reveal that several convolutional architectures effectively address the classification problem using transfer learning. DenseNet stands out as a particularly promising option. Future investigations should explore its performance in scenarios closer to real-world clinical diagnostics. Such studies would benefit from collaboration with medical professionals, aiming to develop a broader image database and align the structural design with the conditions encountered by frontline healthcare workers.

**Acknowledgments**

The authors extend their gratitude to CNPq for the financial support.

**References**

1. X. Zhang, Y. Tan, Y. Ling, G. Lu, F. Liu, Z. Yi, X. Jia, M. Wu, B. Shi, S. Xu et al. "Viral and host factors related to the clinical outcome of COVID-19." Nature, vol. 583, no. 7816, pp. 437–440, 2020.
2. Ministry of Health of Brazil. "Clinical and Laboratory Diagnosis" (Accessed: 2020-11-05). Available at: https://coronavirus.saude.gov.br/diagnostico-clinico-e-laboratorial.
3. S. Bhattacharya, P. K. Reddy Maddikunta, Q.-V. Pham, T. R. Gadekallu, S. R. Krishnan S, C. L. Chowdhary, M. Alazab, and M. Jalil Piran. "Deep learning and medical image processing for coronavirus (COVID-19) pandemic: A survey." Sustainable Cities and Society, vol. 65, pp. 102589, 2021.
4. S. Basu, S. Mitra, and N. Saha. "Deep Learning for Screening COVID-19 using Chest X-Ray Images," 2020.
5. R. Jain, M. Gupta, S. Taneja, and D. J. Hemanth. "Deep learning-based detection and analysis of COVID-19 on chest X-ray images." Applied Intelligence, vol. 51, no. 3, pp. 1690–1700, 2021.
6. T. Rahmann. "COVID-19 Radiography Database on Kaggle" (Accessed: 2020-11-05). Available at: https://www.kaggle.com/tawsifurrahman/covid19-radiography-database.
7. S. J. Pan and Q. Yang. "A Survey on Transfer Learning." IEEE Transactions on Knowledge and Data Engineering, vol. 22, no. 10, pp. 1345–1359, 2010.

8. O. Russakovsky, J. Deng, H. Su, J. Krause, S. Satheesh, S. Ma, Z. Huang, A. Karpathy, A. Khosla, M. Bernstein et al. "Imagenet large scale visual recognition challenge." International Journal of Computer Vision, vol. 115, no. 3, pp. 211–252, 2015.
9. ImageNet Project. "ImageNet Overview" (Accessed: 2020-11-05). Available at: http://www.image-net.org/about-overview.
10. I. Goodfellow, Y. Bengio, and A. Courville. Deep Learning. MIT Press, 2016. Available at: http://www.deeplearningbook.org.
11. J. P. Cohen, P. Morrison, L. Dao, K. Roth, T. Q. Duong, and M. Ghassemi. "COVID-19 image data collection: Prospective predictions are the future" (Accessed: 2020-11-05). Available at: https://github.com/ieee8023/covid-chestxray-dataset.
12. Torch Contributors. "Optimizers Used and Their Differences" (Accessed: 2020-11-05). Available at: https://pytorch.org/docs/stable/optim.html.
13. A. Krizhevsky, I. Sutskever, and G. E. Hinton. "Imagenet classification with deep convolutional neural networks." Communications of the ACM, vol. 60, no. 6, pp. 84–90, 2017.
14. G. Huang, Z. Liu, L. Van Der Maaten, and K. Q. Weinberger. "Densely Connected Convolutional Networks." In Proceedings of the IEEE Conference on Computer Vision and Pattern Recognition (CVPR), pp. 2261–2269, 2017.
15. C. Szegedy, W. Liu, Y. Jia, P. Sermanet, S. Reed, D. Anguelov, D. Erhan, V. Vanhoucke, and A. Rabinovich. "Going deeper with convolutions." In Proceedings of the IEEE Conference on Computer Vision and Pattern Recognition (CVPR), pp. 1–9, 2015.
16. M. Sandler, A. Howard, M. Zhu, A. Zhmoginov, and L.-C. Chen. "Mobilenetv2: Inverted residuals and linear bottlenecks." In Proceedings of the IEEE Conference on Computer Vision and Pattern Recognition, pp. 4510–4520, 2018.
17. K. He, X. Zhang, S. Ren, and J. Sun. "Deep Residual Learning for Image Recognition." In Proceedings of the IEEE Conference on Computer Vision and Pattern Recognition (CVPR), pp. 770–778, 2016.
18. X. Zhang, X. Zhou, M. Lin, and J. Sun. "Shufflenet: An extremely efficient convolutional neural network for mobile devices." In Proceedings of the IEEE Conference on Computer Vision and Pattern Recognition, pp. 6848–6856, 2018.
19. F. N. Iandola, S. Han, M. W. Moskewicz, K. Ashraf, W. J. Dally, and K. Keutzer. "SqueezeNet: AlexNet-level accuracy with 50x fewer parameters and < 0.5 MB model size." arXiv preprint, arXiv:1602.07360, 2016.
20. S. Liu and W. Deng. "Very deep convolutional neural network-based image classification using small training sample size." In Proceedings of the 2015 3rd IAPR Asian Conference on Pattern Recognition (ACPR), pp. 730–734, 2015.
21. ImageNet Project. "The ImageNet Large Scale Visual Recognition Challenge (ILSVRC)" (Accessed: 2020-11-05). Available at: http://www.image-net.org/challenges/LSVRC/.
22. Radiology Assistant - Educational Site of the Radiological Society of the Netherlands. "Images of a young patient with pneumonia" (Accessed: 2021-04-15). Available at: https://radiologyassistant.nl/chest/chest-x-ray.
23. Radiology Assistant - Educational Site of the Radiological Society of the Netherlands. "An 83-year-old male with mitral insufficiency and pulmonary hypertension diagnosed with COVID-19 infection" (Accessed: 2021-04-15). Available at: https://radiologyassistant.nl/chest/covid-19/covid19-imaging-findings.